\documentclass[runningheads]{llncs}
\usepackage[T1]{fontenc}
\usepackage{graphicx}
\usepackage{hyperref}
\usepackage{color}

\begin{document}
\title{Analyzing Concentration, Temporal Routines and Targeting in Public Ransomware Leak Site Data}
\titlerunning{Analyzing Ransomware Leak Site Data}

\author{Lea Müller\orcidID{0009-0009-6538-8046} \and
York Yannikos\orcidID{0009-0001-2751-5253}}

\authorrunning{L. Müller, Y. Yannikos}

\institute{Fraunhofer Institute for Secure Information Technology (SIT) \\
National Research Center for Applied Cybersecurity (ATHENE), Rheinstr. 75, 64295 Darmstadt, Germany\\
\email{\{lea.mueller,york.yannikos\}@sit.fraunhofer.de}}

\maketitle

\begin{abstract}
Ransomware has grown to become one of the most damaging types of cybercrime, affecting private and public organizations in any sector. While early types of ransomware targeted many victims via automated attacks, ransomware groups have started to specifically target organizations and companies in the expectation of receiving larger ransoms. To increase the pressure on victims, most groups host so-called data leak sites, where information about their victims is made public. The shift towards ‘human-operated’ ransomware together with easily accessible behavioral traces available from data leak sites makes research investigating operational regularities of ransomware groups of interest. Using leak site posts as behavioral traces of ransomware groups, we created a dataset consisting of over 27,000 posts from 325 groups. Based on this dataset, we analyzed victim concentration, temporal routines and targeting regularities. Our findings suggest that groups do not behave entirely random. Instead, the observable traces found on leak sites show concentration of activity, temporal routines and selective patterns.

\end{abstract}

\section{Introduction}
Ransomware has grown to become one of the most financially damaging types of cybercrime over the past decade, with ransom payments amounting to USD 820 million in 2025 alone~\cite{chainalysis_team_crypto_2026}; a figure that does not even take into account the damage incurred through loss of business or reputation. Affecting private and public organizations in any sector~\cite{oz_survey_2022}, ransomware is described as a key threat by both industry~\cite{sophos_sophos_2024} and public authorities~\cite{european_union_agency_for_cybersecurity_enisa_enisa_2025,internet_crime_complaint_center_internet_2025}. For example, according to the European Union Agency for Cybersecurity (ENISA)~\cite{european_union_agency_for_cybersecurity_enisa_enisa_2025}, ransomware attacks and data breaches account for over 95\% of cybercriminal activities affecting EU organizations, with the latter directly resulting from the former. Europol describes ransomware as a key threat in the European Union and observes a steady increase in attacks~\cite{europol_evolving_2026}. While the majority of ransomware attacks affect small and medium-sized enterprises, critical infrastructure and governments have also suffered significant incidents in the past~\cite{whelan_reconceptualising_2024}.

Ransomware is a type of malware that encrypts or otherwise blocks access to systems or data~\cite{beaman_ransomware_2021,oz_survey_2022} in order to extort payment from a victim~\cite{oz_survey_2022}. In the past, ransomware targeted many victims via automated attacks~\cite{okane_evolution_2018}, demanding relatively low ransom amounts. With the rise of `big game hunting' and Ransomware-as-a-Service (RaaS), ransomware groups started to more specifically target high-value organizations. With RaaS, cybercriminals can obtain ransomware from operators to execute attacks, eliminating the need to create their own malware. This enables criminals with minimal programming skills to orchestrate ransomware attacks and earn money through extortion~\cite{meland_ransomware-as--service_2020}. RaaS structures often involve a core group developing the malware and coordinating operations, and so-called affiliates performing attacks and sharing profit with the core group~\cite{whelan_reconceptualising_2024}. `Big game hunting' describes targeted ransomware attacks, whereby groups target large companies and organizations rather than individuals in the expectation of receiving larger payments~\cite{beaman_ransomware_2021,matthijsse_your_2023,meland_ransomware-as--service_2020,oz_survey_2022}. Since targeting specific, high-prospect organizations has increased the resources cybercriminals invest, a shift in extortion tactics has been observed. From 2019 onward, ransomware groups have started to host so-called data leak sites to increase pressure on their victims~\cite{german_federal_office_for_information_security_bsi_lage_2022,raj_modern_2024}. In 2020, this procedure became a trend, with more and more ransomware groups using data leak sites as an additional lever of extortion~\cite{german_federal_office_for_information_security_bsi_lage_2022}. Data leak sites are dark web platforms operated by ransomware groups. These sites are used to announce new victims and often include a deadline for them to pay a ransom to prevent their data from being published. While early forms of ransomware simply encrypted a victim's data and demanded a ransom for decryption, extortion now focuses on coercing victims to pay for their data to not be released publicly~\cite{europol_evolving_2026}.

The trend of targeting specific organizations has led perpetrators to manually perform attacks instead of using autonomous malware~\cite{oz_survey_2022,rauf_ali_khan_sequence_2026}. This shift towards `human-operated' ransomware – in contrast to earlier ransomware families spreading autonomously~\cite{microsoft_threat_intelligence_human-operated_2020} – has made research investigating operational regularities of ransomware groups of interest. If attacks are performed by humans, routines or targeting preferences may be observed by analyzing ransomware groups' behavior. While the proliferation of big game hunting has shifted ransomware groups' focus on companies and organizations, little is known about other factors that may influence temporal routines or targeting preferences, such as working conditions or victim characteristics.

Most research in the field of ransomware focuses on its evolution (e.g.,~\cite{okane_evolution_2018}) or approaches to detection and defense (e.g.,~\cite{beaman_ransomware_2021,oz_survey_2022,rauf_ali_khan_sequence_2026}); however, ransomware groups' operational behavior, like temporal routines or targeting, is rarely analyzed. Existing research often has a narrow focus, with analyses concentrating on either a select set of nations or a select set of ransomware groups. Whelan et al.~\cite{whelan_analysing_2025} analyze ransomware attacks in Australia, Canada, New Zealand and UK from 2020 to 2022. Their dataset includes 865 ransomware incidents across the four nations. Data was obtained from a security company that collects information from ransomware groups' leak sites. They find that different ransomware groups target different countries with varying prevalence, suggesting targeting preferences. They also observe that some sectors are targeted more often, with the sector Industrials being the most prevalently attacked while other sectors, such as education or energy, experience fewer attacks. Kim et al.~\cite{kim_cryptocurrency-driven_2025} analyze 20 ransomware groups active in the Arab world, based on 226 incidents disclosed by these groups. Their analysis focuses on groups with a high level of activity in 22 Arab countries, examining incidents reported between 2020 and 2023. In terms of temporal routines, the authors find that incidents do not appear to be connected to specific times or seasons. They also find that the main sectors targeted by ransomware groups are commercial facilities and critical manufacturing. Phipps and Nurse~\cite{phipps_inside_2026} analyze the three ransomware groups Conti, LockBit and BlackCat/ALPHV, based on articles obtained through web search. Their analysis focuses on groups' `origins, structure, organisation, dynamics and nature', arguing that these areas are key characteristics of ransomware groups. In terms of global targeting, they observe opportunistic behavior, with no specific focus on or exclusion of certain sectors or nations; with an exception of nations linked to the former Soviet Union.

Prior studies investigating the targeting behavior of ransomware groups often have a narrow scope, either by concentrating on a select set of ransomware groups (e.g.,~\cite{phipps_inside_2026}) or a select set of nations (e.g.,~\cite{kim_cryptocurrency-driven_2025,whelan_analysing_2025}). Additionally, sectors or nations are often analyzed based on raw incident counts (e.g.,~\cite{aggarwal_ransomware_2023,kim_cryptocurrency-driven_2025,whelan_analysing_2025}) which neglects that incident counts may be dependent on factors such as a sector's or nation's size or economic scale. For example, the United States are consistently named as the nation most affected by ransomware attacks, but are also the nation with the highest economic scale worldwide~\cite{world_bank_group_gdp_nodate}.

To address the gaps identified in prior research, we investigate a large dataset of ransomware leak posts without deliberately excluding certain ransomware groups or nations. Since most ransomware groups operate data leak sites, which are inherently intended as broadcast channels to reach a wide audience and increase pressure on victims, data about ransomware groups' posting behavior can easily be collected. Treating leak site posts as behavioral traces, this study aims to analyze whether ransomware groups exhibit behavioral regularities across organizational, temporal, geographic and sectoral dimensions. To this end, we create a dataset consisting of over 27,000 leak site posts from 325 groups by aggregating information of two open-source ransomware monitoring tools, RansomLook~\cite{ransomlook_ransomware_nodate} and Ransomware.live~\cite{ransomwarelive_ransomware_nodate}. We then provide an exploratory empirical characterization of ransomware groups' leak site posting behavior using this dataset. We analyze victim concentration across groups, temporal routines as well as target selection. The objective of this article is to understand whether ransomware groups employ opportunistic behavior, similar to early-day ransomware using automatic targeting, or whether groups show non-random characteristics in their operations.

Our analyses are structured around four types of findings: 1) Ecosystem structure, analyzing the concentration of activity across ransomware groups. 2) Temporal routines, analyzing whether temporal patterns can be observed in leak post behavior. 3) Geographic target-selection regularities, analyzing whether certain nations are over-/underrepresented relative to their economic exposure. 4) Sectoral target-selection regularities, analyzing whether certain sectors are over-/underrepresented relative to a market-sector baseline.

We make the following contributions:
\begin{itemize}
    \item Analysis of a curated dataset compiled from two open-source ransomware monitoring tools. Our dataset includes 27,629 leak posts from 325 ransomware groups.
    \item Based on analysis of victim concentration, we offer evidence for strong concentration of observable activity on only a few highly active groups.
    \item Based on analysis of temporal routines, we offer evidence for concentration of posting activity on weekdays. Analysis also indicates a seasonal pattern with higher activity in Q4 and lower activity in January.
    \item We compare geographic and sectoral incident distributions with a baseline of economic scale to identify nations and sectors that are over- or underrepresented in ransomware attacks.
\end{itemize}

Overall, while ransomware groups' behavior is often observed to be unstable~\cite{europol_evolving_2026}, our findings suggest that groups do not behave entirely random. Instead, the observable traces found on data leak sites indicate concentration of activity, temporal routines and selective patterns.

The rest of this paper is structured as follows: Section~\ref{sec:methods} presents the methodology, including data collection, curation of our leak post dataset, and the analytical approaches employed. Section~\ref{sec:results} presents the results, structured around activity concentration, temporal routines, and target-selection regularities. Section~\ref{sec:discussion} discusses our observations and the work's limitations.

\section{Methods}\label{sec:methods}
\subsection{Data Collection and Data Curation}\label{subsec:data}
Data was collected from two open-source ransomware monitoring tools, RansomLook~\cite{ransomlook_ransomware_nodate} and Ransomware.live~\cite{ransomwarelive_ransomware_nodate}. Both services track ransomware groups' posts and activities by monitoring leak sites, providing aggregated information about various active and inactive groups. Additionally, some of the entries on Ransomware.live list the organization's nationality and sector. The first victims listed by RansomLook are dated in 2020. The first victims listed by Ransomware.live are dated in 2013. From both sources, we retrieved data on all active and inactive groups with at least one leak site post. At the time of data collection (March 4, 2026) and after removal of duplicate entries, RansomLook listed 25,761 leak posts from 257 groups and Ransomware.live listed 25,656 leak posts from 298 groups. For each post, the collected data includes: timestamp, name of the ransomware group, post title (usually the name of the organization), nationality of the organization where available, sector of the organization where available.

In a first step, for the purpose of merging the data from the two aforementioned services, ransomware groups' names were brought into a standardized format. As some groups were listed with different names by the two services, names were compared and standardized to merge the data in the next step. After standardizing group names, 325 unique groups remained from both services.

In a next step, the two datasets were merged. To be able to merge the two datasets of over 25,000 entries each, we compared the tuples of organization's name and ransomware group's name in both datasets automatically. Since slight variations in organizations' names were observed between the two datasets, we normalized names in a first step and computed similarity between two names in a second step. The similarity of two names was computed as an equation of $\frac{2m}{t}$, with $m$ being the number of matches and $t$ being the total number of elements in both sequences. This results in a similarity score between 0 (= no similarity) and 1 (= identical sequences). The normalization procedure was improved iteratively by inspecting entries manually and identifying causes of false positives (i.e., two entries that should not be merged receiving a high similarity score) and false negatives (i.e., two entries that should be merged receiving a low similarity score). For example, some ransomware groups add prefixes like `full data leak of' to a leak post, which can have significant but irrelevant impact on the similarity score. Iterative refinement of the normalization procedure allowed us to catch such instances and thus reduce the number of errors. Other normalization steps included turning all characters to lowercase, stripping white spaces, removing prefixes and suffixes, or unescaping HTML entities. The two datasets were then merged based on a comparison of normalized organization names. Two entries were treated as a match if the similarity score was $\geq0.8$. Entries with a similarity score $<0.8$ were treated as disparate entries. Matches with a similarity score $\geq0.8$ and $<1.0$ were checked manually and only if the entries were identical, the match was accepted. Matches with a similarity score of 1.0 were accepted automatically and no manual check was done for these instances. Merging the two datasets resulted in a total of 28,127 entries from 325 groups.

In a last step, the entire dataset was checked manually to remove entries not containing incident information. For example, some entries were announcements to inform site visitors about new channels or collaborations. 498 entries were removed, resulting in a total of 27,629 leak posts. Additionally, we observed 791 instances where victims' names had been anonymized. In most cases, these entries did not provide enough information to identify an organization's nationality or sector.

For the purpose of analyzing geographic preferences in targeting, we employed three rule-based steps to identify organizations' nationality. In the first step, where available, we copied the information about an organization's nationality as provided by Ransomware.live. In the second step, we identified the nationality from the top-level domain (TLD) for the entries for which a domain was given. This was only possible for leak posts where the ransomware group provided the victim's domain. In the third step, we implemented a keyword-based approach to identify leak posts that contained the organization's nationality in their title. Employing these three steps, nationality was identified for 18,041 victims (65.30\% of victims). It should be noted that classification of nationality was not achieved for all victims. Firstly, researching over 9,000 organizations manually is not feasible for the purpose of this study. Secondly, an organization may be international and classification into one nationality not possible. Additionally, 791 of the incidents were anonymized, with the majority of anonymized posts not providing enough information to be able to identify the organization or information about its nationality.

For the purpose of analyzing sectoral preferences in targeting, we first defined a number of sectors to be part of our analysis. We used the 11 Global Industry Classification Standard (GICS) sectors, expanded to include government and education as additional sectors. The GICS is an industry classification framework developed by MSCI~\cite{msci_global_nodate} and S\&P Dow Jones Indices~\cite{sp_global_gics_nodate} and classifies organizations into 11 sectors. Government and education institutions, however, are not covered by this classification. After having defined a set of sectors, we employed two rule-based steps to identify the sector of victim organizations. In the first step, where available, we mapped the sector information provided by Ransomware.live to the 13 sectors selected for our analysis. For the remaining entries, classification was realized via a keyword-based approach. Keywords were defined for the 13 sectors. If the name of an organization contained one of the keywords, it was classified into the respective sector. Sector information was identified for 16,868 victims (61.05\% of victims). It should be noted that classification of the sector was not achieved for all victims. Firstly, researching over 10,000 organizations manually is not feasible for the purpose of this study. Secondly, over 791 organizations' names were anonymized, with the majority of anonymized posts not providing enough information to be able to identify the organization or the sector it operates in.

\subsection{Analyses}
Based on the dataset of leak site posts, we employed explorative data analyses to understand activity concentration across groups, temporal behavior, and target-selection regularities.

\subsubsection{Victim Concentration across Groups}
To analyze victim concentration across groups, we calculated the number of leak site posts per group. To assess concentration, we calculated the concentration ratio $CR_k$ for $k = 0.01n$, $0.05n$, $0.10n$, $0.20n$, with $n=325$ being the number of ransomware groups in our dataset. I.e., we calculated the concentration ratio for the largest 1\%, 5\%, 10\%, and 20\% of groups.

\subsubsection{Temporal Behavior}
We analyzed the distribution of leak site posts over weekdays by calculating weekend-weekday ratio based on raw leak post counts per weekday. We also calculated daily averages of z-standardized ransomware incident counts to identify days with activities above or below average. Z-standardized values were used to control for long-term trends. It should be noted that we only considered incidents between January 2018 and 04 March 2026. Earlier incidents were excluded due to rare occurrence of leak site posts.

We analyzed the distribution of leak site posts over months using annual z-standardized values to control for long-term trends. Due to rare occurrence of leak site posts prior to 2019, we only considered posts between January 2019 and February 2026. 

\subsubsection{Geographic Regularities}
To identify geographic regularities, we first calculated each nation's relative number of ransomware attacks. It should be noted that relative numbers were calculated based on the total number of organizations for which we were able to identify the nationality (see Section~\ref{subsec:data}). This means that we set the figure of 18,041 incidents as 100\% in order to calculate each nation's share, so as to avoid artificially generating lower relative numbers. Since raw percentages fail to capture nations' economic scale, we calculated deviations of observed incident distribution from an economic baseline. Deviations were calculated for all nations that account for a cumulative 95\% of all ransomware incidents (52 nations). As a baseline, we used gross domestic product (GDP) as a proxy for nations' economic scale. Nations' GDP in 2024 was sourced from~\cite{world_bank_group_gdp_nodate}. Since Taiwan is not listed in this database, GDP for Taiwan in 2024 was sourced from~\cite{international_monetary_fund_gdp_nodate}. Deviations from the baseline indicate whether a nation is over- or underrepresented relative to its economic weight.

\subsubsection{Sectoral Regularities}
To identify sectoral regularities, we first calculated each sector's relative number of ransomware attacks. Similar to the analysis of geographic regularities, we set the figure of 16,868 incidents as 100\% (see Section~\ref{subsec:data}). Since raw sectoral distributions may lack informative value, we calculated the deviation of ransomware incident distribution from a baseline representing each sector's economic scale. The sectoral baseline was derived from data about value added to GDP per industry, as published by the U.S. Bureau of Economic Analysis (BEA)~\cite{us_bureau_of_economic_analysis_value_2026}. The industries listed by BEA were mapped to the 13 sectors (eleven GICS sectors, expanded to include government and education) and value added was calculated as a sum of the mapped industries. As the sectoral baseline, we used relative value added by each sector as a proxy for economic scale. We then calculated the deviation of each sector's share of ransomware incidents from the distribution in the sectoral baseline.

\section{Results}\label{sec:results}
\subsection{Victim Concentration across Groups}
\begin{figure}
    \centering
    \includegraphics[width=0.95\textwidth]{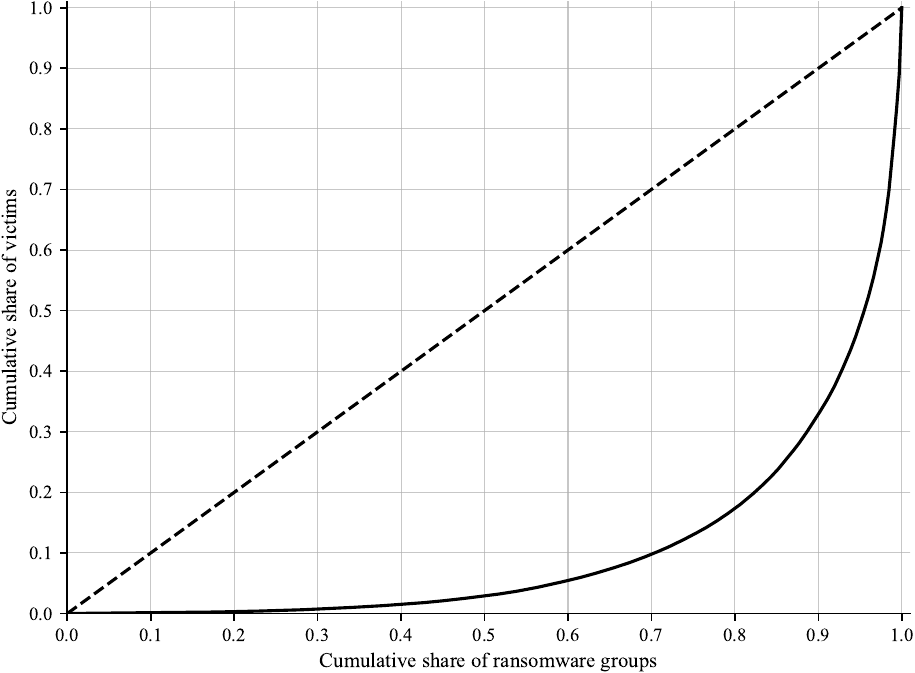}
    \caption{Lorenz curve of victim concentration across ransomware groups}
    \label{fig1}
\end{figure}

The number of leak site posts ranged from 1 for the smallest groups to 3,038 for the biggest group, with a mean of 85.01 leak posts per group. Figure~\ref{fig1} shows the Lorenz curve of the victim concentration across ransomware groups, indicating a substantial concentration in the distribution of incidents. We report the concentration ratio for the largest 1\%, 5\%, 10\%, and 20\% of groups. $CR_{0.01n} = CR_{3} = 21.29\%$; $CR_{0.05n} = CR_{16} = 51.84\%$; $CR_{0.10n} = CR_{33} = 67.45\%$; $CR_{0.20n} = CR_{65} = 82.66\%$. Additionally, we report the share of groups needed to explain 50\%, 75\%, 90\%, and 95\% of attacks. The top 15 groups (4.62\% of groups) account for over 50\% of attacks; the top 46 groups (14.15\%) account for 75\% of attacks; the top 97 groups (29.85\%) account for 90\% of attacks; and the top 135 groups (41.54\%) account for 95\% of attacks. This illustrates that the bottom 190 groups (58.46\%) account for less than 5\% of all observed ransomware incidents.

\subsection{Temporal Behavior}
\begin{figure}
    \includegraphics[width=\textwidth]{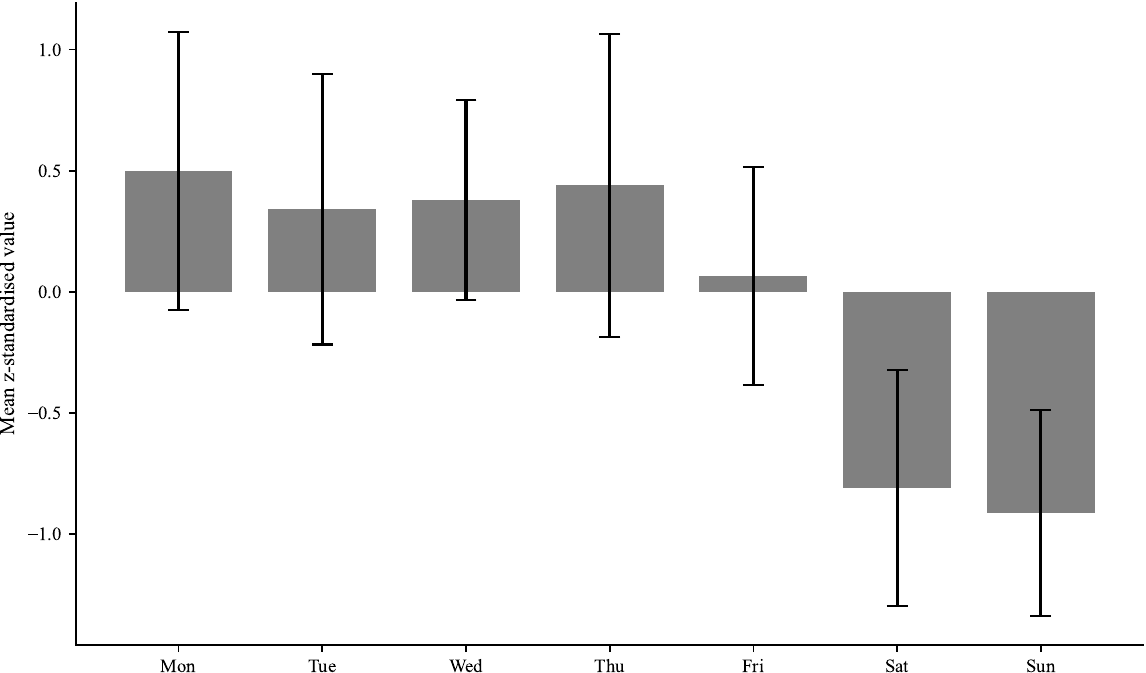}
    \caption{Daily averages of z-standardized ransomware incident counts. Values are z-standardized within each year to control for long-term trends. Positive values indicate above-average activity, while negative values indicate below-average activity. Error bars represent 95\% confidence intervals.}
    \label{fig2}
\end{figure}

We analyzed the distribution of leak site posts over weekdays. Weekend-weekday ratio was $0.51$, meaning that approximately 49\% fewer leaks were posted on weekends than on weekdays. Figure~\ref{fig2} shows the distribution of attacks over weekdays. After z-standardization, significantly lower activity on weekends can be observed. No specific activity patterns can be observed during the working week.

\begin{figure}
    \includegraphics[width=\textwidth]{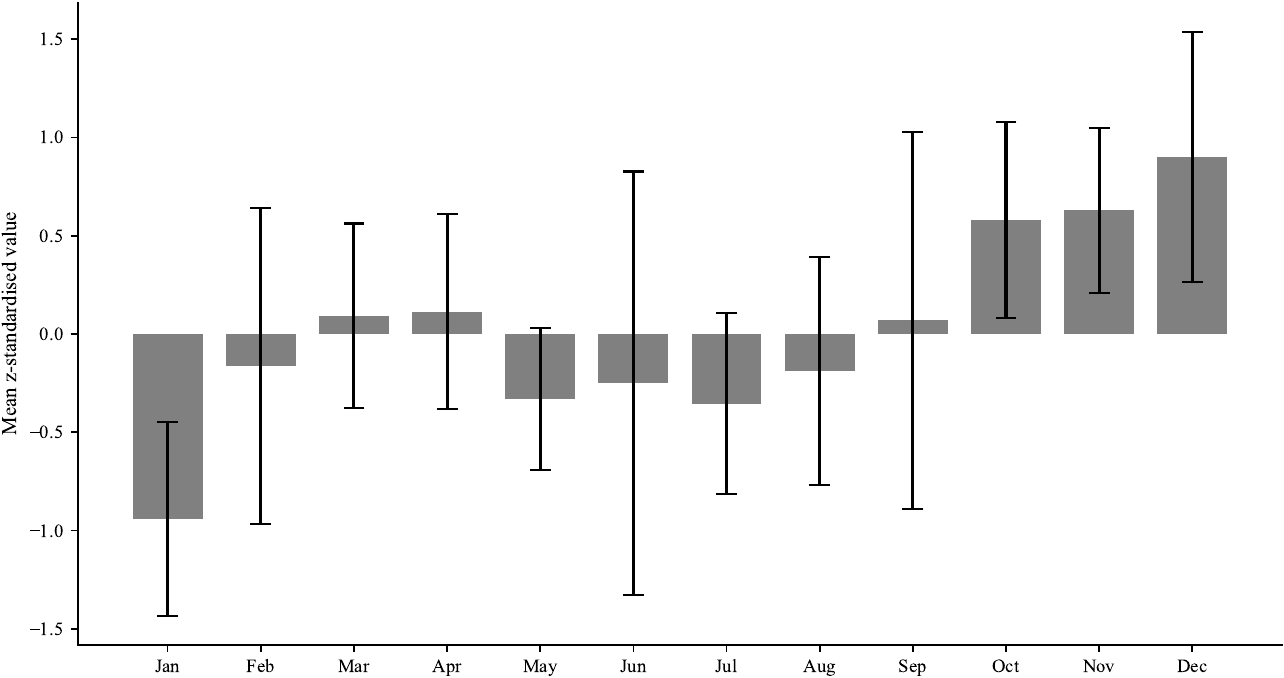}
    \caption{Monthly averages of z-standardized ransomware incident counts. Values are z-standardized within each year to control for long-term trends. Positive values indicate above-average activity relative to the respective yearly mean, while negative values indicate below-average activity. Error bars represent 95\% confidence intervals.}
    \label{fig3}
\end{figure}

Figure~\ref{fig3} shows the distribution of attacks per month. After z-standardization, a pattern of relative overactivity emerges in Q4, particularly in December, with October, November and December showing positive deviations from the yearly mean. January demonstrates a negative deviation from the yearly mean. These findings suggest a possible seasonal component in the attack behavior of ransomware groups. However, the relatively wide error bars indicate substantial inter-annual variability, suggesting that a seasonal effect is present but not consistently pronounced across all years.

\subsection{Geographic Regularities}
\begin{figure}
    \includegraphics[width=\textwidth]{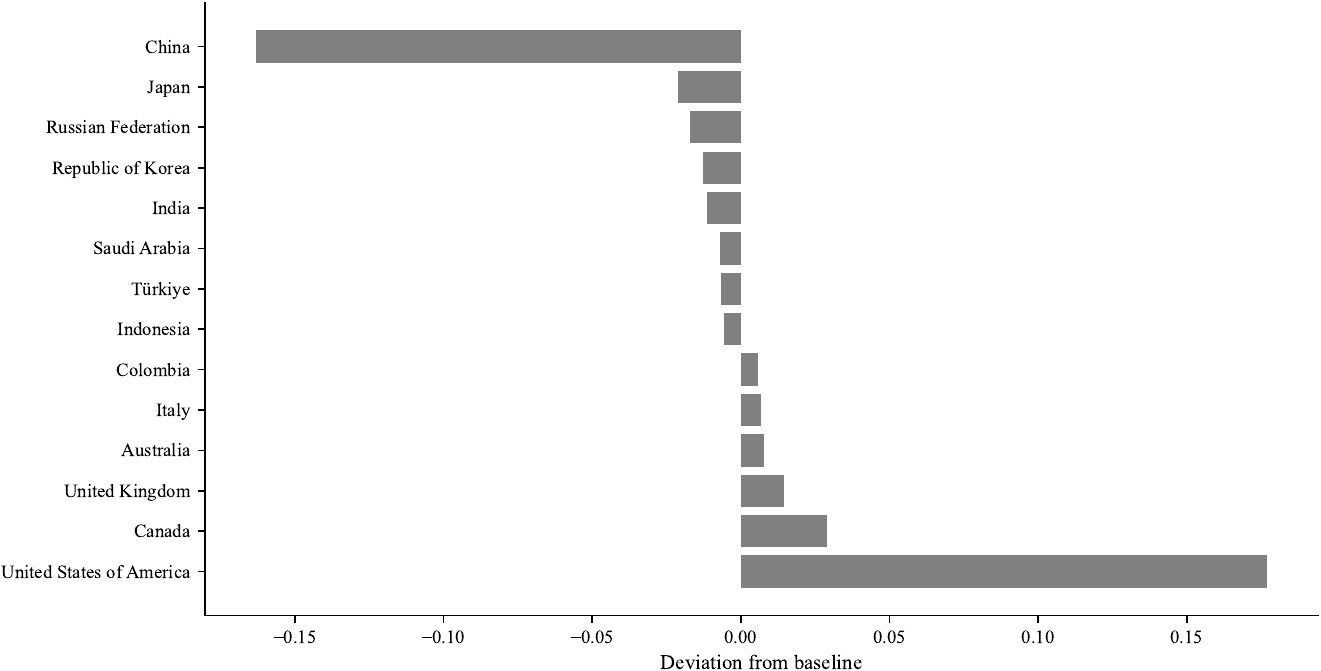}
    \caption{Deviation of ransomware incident distribution from GDP baseline. Deviations illustrate the overrepresentation (positive deviation) and underrepresentation (negative deviation) in ransomware incidents relative to nations' economic scale. Note: Figure only shows nations where deviation exceeds $\pm0.5\%$.}
    \label{fig4}
\end{figure}

Of the incidents for which the nationality of the victim organization was identified, the majority occurred in the USA (43.62\%), followed by Canada (4.93\%), the United Kingdom (4.77\%), and Germany (4.25\%). Since raw percentages fail to capture the economic scale of nations, we report deviations of observed incident distribution from an economic baseline. The deviations are shown in Figure~\ref{fig4}. The highest underrepresentation is observed for China, followed by Japan and Russia; the highest overrepresentation is observed for the United States, followed by Canada and the United Kingdom.

\subsection{Sectoral Regularities}
\begin{figure}
    \includegraphics[width=\textwidth]{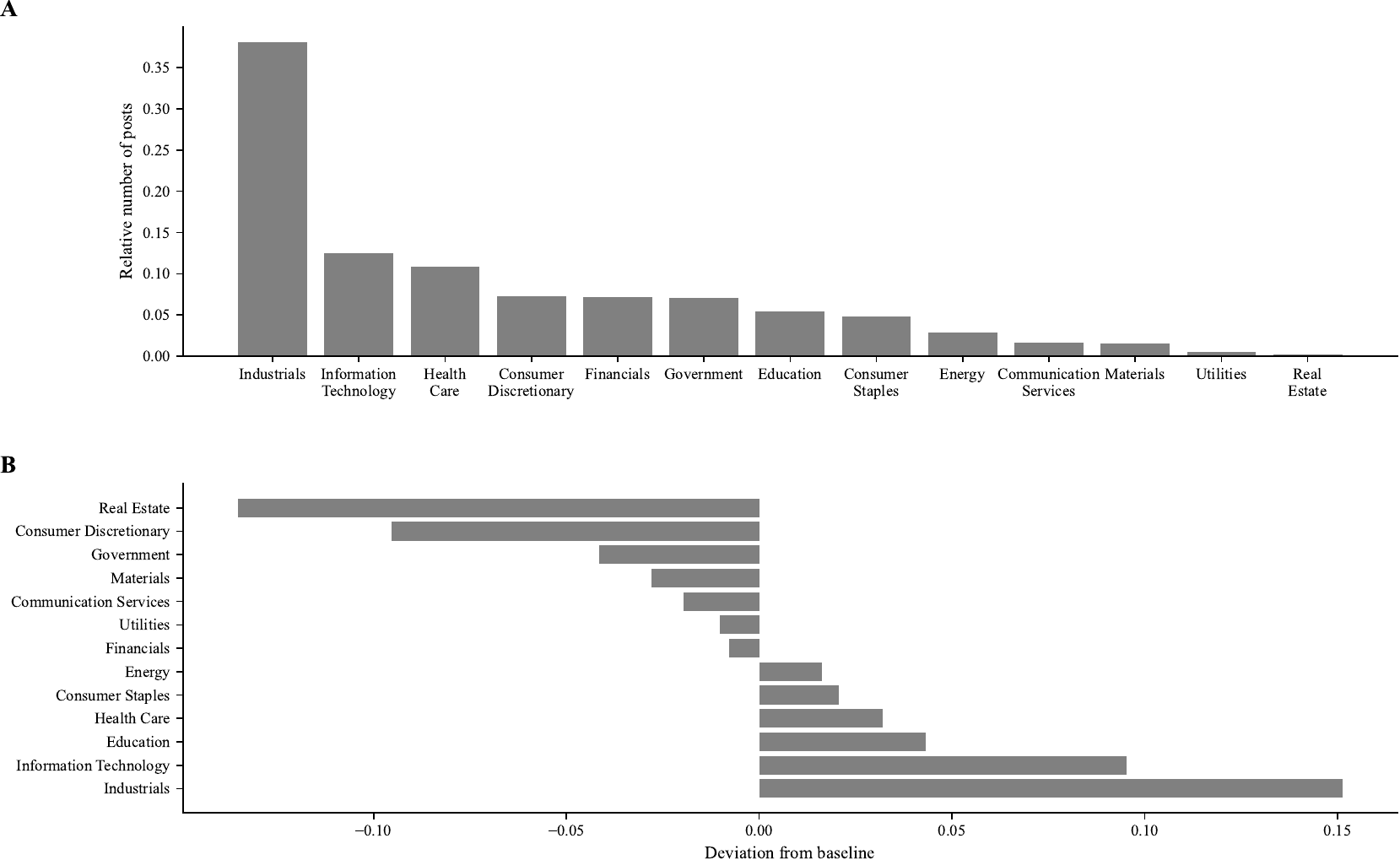}
    \caption{Sectoral distribution of incidents. (A) Relative number of incidents per sector. (B) Deviation of ransomware incident distribution from a sectoral baseline. Deviations illustrate the sectoral overrepresentation (positive deviation) and underrepresentation (negative deviation) in ransomware incidents relative to the baseline distribution.}
    \label{fig5}
\end{figure}

Figure~\ref{fig5} (A) shows the distribution of attacks across 13 sectors. Most attacks targeted Industrials (38.02\%), followed by Information Technology (12.47\%) and Health (10.91\%). Since raw sectoral distributions may lack informative value, we also report the deviation of ransomware incident distribution from a baseline representing each sector's economic scale. The deviations are illustrated in Figure~\ref{fig5} (B). Negative values indicate underrepresentation in incidents and positive values indicate overrepresentation. The strongest overrepresentation is observed for Industrials with 15.12\% deviation, followed by Information Technology (9.52\%) and Education (4.32\%). The strongest underrepresentation is observed for Real Estate with 13.51\% deviation, followed by Consumer Discretionary (9.52\%) and Government (4.16\%).

\section{Discussion}\label{sec:discussion}
We collected and curated a dataset of 27,629 unique leak site posts from 325 ransomware groups by aggregating data from two open-source ransomware monitoring tools~\cite{ransomlook_ransomware_nodate,ransomwarelive_ransomware_nodate}. Based on this dataset, we analyzed victim concentration across groups, temporal routines as well as geographic and sectoral regularities. The objective of our analysis is to understand whether human-operated ransomware reflects opportunistic behavioral patterns or whether non-random characteristics can be observed from behavioral traces on data leak sites.

Analysis of the concentration of victims across groups showed that incident counts are dominated by a small number of highly active groups. In contrast, a large number of less active groups contribute only a small proportion of attacks. This finding suggests that the ransomware ecosystem is not an equally distributed market of many groups but one that is dominated by a few highly visible or active groups. A possible reason for the strong concentration of activity is the method of operation of RaaS. Ransomware groups involved in RaaS are often described as a structure that involves a core group developing the ransomware and coordinating operations, and so-called affiliates performing attacks and sharing profit with the core group~\cite{whelan_reconceptualising_2024}. Since RaaS providers depend on affiliates to extort ransoms, they may be `incentivised to offer convenient all-in-one solutions to attract affiliates and establish a reputable and stable brand-identity'~\cite{europol_evolving_2026}. As a result, some groups may offer more attractive conditions to attract affiliates, such as offering easy-to-use products or granting a larger share of the ransom extorted. For example, Phipps and Nurse~\cite{phipps_inside_2026} observed that ransomware groups demonstrate efforts to increase the ease of use of their services and lower the level of technical skill required to attract affiliates. Cable et al.~\cite{cable_showing_2024} showed that the splits received by affiliates may differ between ransomware groups, with some groups granting a larger share of the ransom and others demanding a larger share for the group itself. It is conceivable that affiliates select the groups they work with based on brand appeal or conditions offered. Groups with higher appeal could thus attract more affiliates and dominate a larger share of the market.

Analysis of temporal routines showed that ransomware groups' activity on data leak sites is concentrated on weekdays. Significantly fewer leaks are posted on weekends. On average, weekends were 49\% less active. One possible explanation for this observation is that ransomware groups exhibit typical organizational working patterns. This finding is consistent with the organizational structure seen in some ransomware groups, such as Conti and REvil. These groups have been described as being structured like legitimate businesses~\cite{paternoster_inside_2025,whelan_internal_2025}, including division of labour and appealing job conditions~\cite{martin_ransomware_2025,whelan_reconceptualising_2024}. For example, an analysis of Conti's human resources practices revealed that most group members were expected to work a standard five-day week~\cite{martin_ransomware_2025}. If similar conditions apply to other groups, the significantly lower number of leak posts on weekends could be a direct result of groups' standard working conditions. It should be noted that, due to the proliferation of targeted ransomware, attacks are usually performed manually~\cite{oz_survey_2022,rauf_ali_khan_sequence_2026}. Therefore, it would be expected to observe some form of human working rhythm, which would not presumably be the case if the malware were operating completely autonomously. However, working hours may not directly result from the job conditions of group members, but rather reflect a greater effectiveness of leak posts on weekdays. As some victimized organizations may not work outside business hours, the observed temporal routines could also reflect an attempt to align with victims' business hours in order to increase the impact of leak posts. For example, Martin et al.~\cite{martin_ransomware_2025} observed that Conti's working hours were presumably aligned with business hours in the United States. It is conceivable that the same applies to a standard five-day week.

Analyzing temporal routines on a monthly level, we observed higher activity in Q4, with December being the month with the most leak posts. We observed lower activity in January. However, considerable inter-annual variability can be observed, suggesting that seasonal routines are present but not consistent across all years. The present analysis cannot determine whether this pattern reflects attacker strategy, operational routines or dataset-specific artifacts.

To examine whether some nations are over- or underrepresented in ransomware incidents, we compared nations' relative number of incidents with a baseline of relative GDP, serving as a proxy for economic scale. Comparison with this baseline revealed that the United States are highly overrepresented in ransomware attacks, followed by Canada, the United Kingdom and Australia. The overrepresentation of some nations could reflect targeting preferences. Particularly the high overrepresentation of the United States, which account for 43.62\% of all attacks in our dataset but only for 25.91\% of global GDP, may reflect a genuine targeting preference of ransomware groups. Considering that the four nations with the highest overrepresentation are English-speaking, preferences may be attributable to language. An analysis of Conti's hiring practices by Martin et al.~\cite{martin_ransomware_2025} revealed that the ransomware group was explicitly looking for new negotiators with `[g]ood knowledge of spoken English'~\cite{martin_ransomware_2025}. While this requirement may reflect English being the standard language in ransomware negotiations~\cite{georgiou_engaging_2026}, it can be assumed that good knowledge of the language of a victim organization may increase the chances of success when moving manually inside a network.

To analyze sectoral over- and underrepresentation, we compared the incident distribution across 13 sectors (11 GICS sectors, expanded to include Government and Education) with a baseline of economic scale. The sector Industrials was observed to be both the sector with the most incidents and the sector with the highest overrepresentation compared to a baseline of economic scale. Activity concentration on the Industrials sector was also observed by Whelan et al.~\cite{whelan_analysing_2025}. While these observations may indicate targeting preferences, the overrepresentation of certain sectors could also suggest lower security levels. Matthijsse et al.~\cite{matthijsse_your_2023} and Phipps and Nurse~\cite{phipps_inside_2026} argue that attacks are based on opportunity rather than dedicated target selection, suggesting that the observed patterns reflect vulnerabilities rather than the preferences of ransomware groups.

The findings of our analyses suggest that ransomware groups demonstrate structured, non-random patterns: a small number of highly active groups dominates the ransomware ecosystem, postings follow typical human working rhythms, and victim distributions deviate from economic baselines across countries and sectors.

\subsection{Limitations}
The contribution of this study should be understood within its methodological and dataset choices. The empirical analyses performed are based on leak site data, meaning that observed patterns do not necessarily reflect actual ransomware incidents. Some incidents may not be posted to leak sites. At the same time, some leak site posts may be false claims made by a ransomware group, meaning that no actual attack is underlying the claim or older posts are reused~\cite{janofsky_ransomware_2024}. As a result, our dataset may not accurately reflect actual attack behavior, as it may exclude some incidents while including false claims. However, leak posts provide publicly observable evidence of the ransomware ecosystem and offer easily accessible data for analysis. Since data leak sites have become prevalent among ransomware groups since 2019~\cite{german_federal_office_for_information_security_bsi_lage_2022}, it is reasonable to assume that most attacks will be publicized on these sites. As a result of our dataset choice, the observations in this study should be interpreted as reflecting disclosure behavior rather than attack behavior. Intrusion and data exfiltration will occur prior to an attack being posted, and attacks may not be posted immediately. The observed temporal routines do not allow conclusions to be drawn about attack behavior.

The dataset used in this study was created by collecting and curating data from two open-source ransomware monitoring tools~\cite{ransomlook_ransomware_nodate,ransomwarelive_ransomware_nodate}. Although this allowed us to create a comprehensive dataset of 27,629 ransomware leak site posts from 325 groups, further leak sites could exist that our data did not include. We tried to reduce this risk by combining data from two services, which meant that we could include information that one of the services might have missed. As both tools provide data on inactive groups, we were also able to consider their behavior, resulting in a large, comprehensive dataset.

To remove duplicates from the dataset, we implemented a process of automatic deduplication. Manual deduplication was considered infeasible for merging the two datasets of 25,761 and 25,656 leak posts. We calculated similarities between two entries, which could lead to both false positives and false negatives. Regarding false positives, non-identical entries could have been merged if the names of two organizations are highly similar or if pre- and suffixes distort the similarity score. Regarding false negatives, identical entries could have been missed by automatic deduplication if one source uses pre- or suffixes or different characters. We tried to reduce errors produced by this approach by implementing a normalization procedure and checking all entries with a similarity score $\geq0.8$ and $<1.0$. Only matches with a similarity score of 1.0 (indicating identical sequences) were merged without a manual check. While this approach reduces false positive matches, false negatives may still occur if the similarity score is lower than 0.8.

For the purpose of identifying the nationality and sector of victims, we implemented a rule-based approach. An inherent constraint of such an approach is that classification is only possible if one of the defined rules applies. We were able to identify the nationality for 65.30\% of organizations and the sector for 61.05\%. Therefore, our analyses of geographic and sectoral regularities are based on subsets of the collected data. Additional constraints arise from the rules applied. Firstly, nationality identification based on top-level domain can produce false results in some cases. For example, \textit{.ai} or \textit{.io} are the TLDs of Anguilla and the British Indian Ocean Territory but are widely used domains outside these territories. At the same time, many organizations use TLDs that are not nation-specific, such as \textit{.com} or \textit{.org}. Nevertheless, using TLDs is a fast and scalable approach that enables the automated identification of victims' nationality for large datasets. Secondly, naming conventions within specific sectors may influence the sector classification used in this study. Since a keyword-based approach was implemented, sectors in which the use of typical keywords in organization names is commonplace may be overrepresented in the subset used for our analyses.

As raw incident counts may not accurately reflect the size or exposure to ransomware threats of a nation or sector, we defined baselines to enable a comparison of observed incident counts with the economic scale of nations and sectors. Using GDP as a proxy for ransomware exposure presents a limitation, since it does not capture either the number of potential targets or their level of protection against cybercrime threats. Other proxies may be conceivable, such as the number of organizations in a country or sector, or organizations' level of security. However, data on the number of organizations in a country or sector is not as easily accessible as data on GDP. Our analysis does not assume that GDP is the only determinant of ransomware targeting. Rather, it uses GDP as a simple baseline of economic size against which observed attack distribution can be compared. Additionally, the sectoral baseline is derived from data on value added by industry, as reported by the U.S. Bureau of Economic Analysis. Consequently, it only represents US companies. The global economic landscape may not follow this distribution. The baselines used in our analyses should be treated carefully, as they represent proxies for economic scale rather than a ground truth for ransomware exposure.

\subsection{Conclusion}
Overall, our findings suggest that public ransomware leak site data does not merely document isolated incidents but reveals behavioral regularities of ransomware groups. Across more than 27,000 leak site posts from 325 groups, visible ransomware activity is highly concentrated among few groups, follows typical human working rhythms, and exhibits deviations from economic baselines across countries and sectors. These findings suggest that leak posts can serve as behavioral traces of human-operated ransomware to analyze underlying operational structures and regularities of the ransomware ecosystem.

\begin{credits}
\subsubsection{\ackname}
This research was supported through the National Research Center for Applied Cybersecurity ATHENE. ATHENE is funded by the German Federal Ministry of Research, Technology and Space and the Hessian Ministry of Science and Research, Arts and Culture.
\end{credits}

\bibliographystyle{splncs04}
\bibliography{bibliography}

\end{document}